\documentclass[prl,preprint,showpacs,preprintnumbers,amsmath,amssymb]{revtex4}

\usepackage{graphicx,color}
\usepackage{dcolumn}
\usepackage{bm}

\begin{document}


\title{Ferromagnetism in $2p$ Light Element-Doped II-oxide and III-nitride Semiconductors}

\author{L. Shen, R. Q. Wu, H. Pan, G. W. Peng, M. Yang, Z. D. Sha and Y. P. Feng}
\email{phyfyp@nus.edu.sg}
 \affiliation{Department of Physics, Faculty of Science, National University of Singapore, Singapore 117542}



\date{\today}

\begin{abstract}
II-oxide and III-nitride semiconductors doped by nonmagnetic 2{\em p} light
elements are investigated as potential dilute magnetic semiconductors
(DMS).
Based on our first-principle calculations, nitrogen doped ZnO, carbon doped
ZnO, and carbon doped AlN are predicted to be ferromagnetic. The ferromagnetism
of such DMS materials can be attributed to a $p$-$d$ exchange-like $p$-$p$
coupling interaction which is derived from the similar symmetry and wave
function between the impurity ($p$-like $t_2$) and valence ($p$) states. We
also propose a co-doping mechanism, using beryllium and nitrogen as dopants in
ZnO, to enhance the ferromagnetic coupling and to increase the solubility and
activity.

\end{abstract}

\maketitle

Magnetic 3$d$ transition metal (3$d$~TM) doped dilute magnetic semiconductors
(DMS) based on III-V and II-VI hosts have been extensively studied due to their
potential applications in spintronics \cite{Dietl.Science}. However, despite
considerable theoretical and experimental efforts, the origin of ferromagnetism
observed in some DMS is still under debate, and whether the ferromagnetic (FM)
property arises from the doped magnetic TM ions which were observed to form
magnetic clusters or secondary phases is still not clear. These
extrinsic magnetic behaviors are undesirable for practical
applications. A possible way to avoid problem related to magnetic precipitate
is to dope semiconductors or oxides with nonmagnetic elements instead of
magnetic TM. Following this idea, room temperature ferromagnetism has been
demonstrated in Cu and Ti doped GaN \cite{wu.GaN:Cu, Lee.GaN:Cu} and ZnO
\cite{Ye.ZnO:Cu, Buchholz.ZnO:Cu}. Besides cationic impurities, theoretical
studies predicted that anion substitutions in IIA-oxide can also lead to
ferromagnetism \cite{kenmochi, elfimov}. This represents a new concept in the
search for potentially useful DMS. Work done so far along this line have been
confined to IIA-oxides, and the mechanism for ferromagnetism in such
anion doped magnetic materials is still not understood. The existing theories
of DMS cannot be applied because they are based on $d$ and $f$ orbitals but
there are no such orbitals in these anion doped materials. There are several
important differences between the 2$p$ and the 3$d$ orbitals which determine
the different  magnetic propoerties of DMS doped with 2$p$~LE (anion) and
3$d$~TM (cation). Firstly, the anion 2$p$ bands of the light element are
usually full in ionic states, leaving no room for unpaired spins compared to
3$d$ bands of TM. Secondly, the spin-orbit interaction of $p$ states is
considerably reduced compared to that of $d$ states since it scales with the
fourth power of the atomic number. Consequently, spin relaxation of DMS doped
with 2$p$ light elements is expected to be suppressed by up to two orders of
magnitude in comparison with 3$d$ cation doped DMS \cite{jacs}. Thirdly,
valence electrons in $p$ states are more delocalized than those in $d$ or $f$
states and have much larger spatial extensions which could promote long-range
exchange interactions. Therefore, Despite suffering from low
solubility\cite{fons, li01, pan}, DMS doped with 2$p$ light elements can be
weak ferromagnets in a highly ordered and low doping concentration.

In this letter, we present a possible pathway toward novel II-oxide and
III-nitride DMS based on doping by 2$p$~LE and explore the origin of their
ferromagnetism. Wide band gap II-oxide and III-nitride semiconductors, such as
ZnO and AlN, which have been widely investigated both theoretically and
experimentally for applications in optoelectronics, are potential host
materials for DMS. Results of our first-principles calculations predict that
nitrogen doped ZnO (ZnO:N), carbon doped ZnO (ZnO:C), and carbon doped AlN
(AlN:C) are ferromagnetic. Surprisingly, almost all magnetic moments reside on
the $p$-states of anions (C, N, or O). Compared to the conventional 3$d$~TM
cation doped DMS, these 2$p$~LE anion doped DMS can provide a possible way to
resolve the clustering problem of magnetic elements and will open another
avenue for producing potentially useful DMS. Since these DMS typically have a low doping
concentration but enough mobile carriers, and only 2$p$ electrons of dopants
and hosts contribute to the magnetism, the origin of ferromagnetism in these
materials challenges our current understanding of ferromagnetism of DMS.
Therefore, a comprehensive understanding of the physics of 2$p$~LE anion doped
DMS is essential for further exploration of new DMS. We provide a physical
insight into the ferromagnetic mechanism of such DMS.

In addition, we propose a co-doping mechanism to enhance the ferromagnetism of
2$p$~LE doped DMS materials. Such an approach has been used to solve the
asymmetry doping problem of ZnO. For example, a co-doping approach
\cite{katayama} was demonstrated to be effective in enhancing the N solubility
and reducing the ionization energy of ZnO:N-based materials, such as ZnO:N+Ga
and ZnO:N+Be \cite{yan,lu-Li+N}. Recently, it was found that ferromagnetism in
DMS can be enhanced by co-doping. For example, Sluiter {\em et al.} used this
approach to promote the FM state and increase the Curie temperature ($T_c$) of
ZnO-based ferromagnets \cite{sluiter}. Based on results of our first-principles
calculations, we predict that FM coupling in ZnO:N can be enhanced by co-doping
with Be, and for the first time, we propose that N doped ZnO has possible
application in spintronics, besides optoelectronics which has been studied
extensively.

We performed band structure and total-energy calculations using
first-principles method based on the density functional theory. The
calculations were carried out using the VASP program \cite{VASP}, with the
generalized gradient approximation \cite{GGA-PBE}, and projector augmented wave
method. The energy cutoff was 400~eV and a $k$-mesh of 2$\times$2$\times$2 was
used for a 108-atom ZnO host supercell (3{\em a}$\times$3{\em a}$\times$3{\em
c}). All atoms were allowed to relax until the Hellmann-Feynman forces acting
on them become less than 0.02~eV/\AA.

We consider first a single anion-substitution (N$_{\rm O}$) which is modeled by
replacing an oxygen atom in the supercell by a nitrogen atom. This corresponds
to a doping concentration of 1.85~at.\%, comparable to experimental doping
level. Based on the calculated total-energy, N$_{\rm O}$ favors a
spin-polarized state and its total energy is 23~meV lower than that of the
non-spin-polarized state. Each N dopant introduces a total magnetic moment of
1.0~$\mu_{B}$ ($\sim $\,$0.4 \mu_{B}$ from N itself, $\sim$\,$0.4\mu_{B}$ from
its twelve second neighboring O atoms, and $\sim$\,$0.1 \mu_{B}$ from its four
nearest neighboring Zn atoms). Fig.~1(a) and 1(b) show the calculated total
density of states (DOS) and partial density of states (PDOS) of the 2$p$ states
of nitrogen and a second nearest neighboring oxigen, respectively. As can be
seen, the N 2$p$ states overlap significantly with those of O 2$p$ near the
Fermi level, suggesting a strong interaction between them. This strong
interaction results in the splitting of the energy levels near the Fermi
energy. The spin-up states are fully occupied but spin-down states are
partially filled. The corresponding spin density distribution is shown in
Fig.~2(a). Different from 3$d$~TM-doping in wurtzite host semiconductors where
the spin is exclusively localized in a tetrahedron formed by the four nearest
neighboring anions of an impurity 3$d$~cation\cite{wu.GaN:Cu}, much of the spin
density in the N doped ZnO are localized on the dopant itself and its twelve
second neighboring O anions, with a minor contribution from the nearest
neighboring Zn atoms. Therefore, the magnetic moment in N doped semiconductors
is mainly contributed by the anions, and it results mainly from the delocalized
2$p$ orbitals. The large spatial extension of the $2p$ states can be clearly
seen in Fig.~2(b).

To investigate the magnetic coupling between N impurities, a pair of N atoms
are incorporated into the same ZnO supercell by substituting two O atoms which
are separated by 3.249~\AA, 6.136~\AA, and 9.252~\AA, respectively.
This corresponds to a doping concentration of 3.7 at.\%.
Results of our calculations show that the magnetic moments of the two N dopants
favor FM coupling in each of the three configurations and the energy of the FM state is
7~meV, 22~meV, and 10~meV lower than that of the corresponding antiferromagnetic (AFM)
state, respectively. This indicates that ZnO:N is a weak ferromagnet in a low
nitrogen concentration. Fig.~3 shows the magnetic coupling between the two N ions
separated by 6.136~\AA. As can be seen, charge carriers localized around the
anions between the N ions are polarized and have the same spin orientation
as that of the N ions. Consequently, these polarized charge carriers mediate
the long-range ferromagnetic coupling between the N ions.

Similar spin polartization was also found in C doped AlN and ZnO. Fig.~1(c) and
1(d) show the total DOS of AlN:C and ZnO:C. The calculated magnetic moments per
carbon dopant are 1.0 and 2.0~$\mu_{B}$ in AlN:C and ZnO:C, respectively. The
spins also mainly reside on the C dopant and its neighboring N or O atoms. Both
systems energetically favor the FM ground state. High Curie temperature (over
400 K) ferromagnetism in C doped ZnO has been demonstrated recently\cite{pan}.

Although no theoretical consensus on the mechanism of ferromagnetism in DMS has
been reached, it is now well established that spin-polarized carriers (holes or
electrons) are crucial for the ferromagnetism in DMS, either free or localized.
Such an understanding of FM origin can be based on band structure models
\cite{Dietl.Science, Akai.PRL, Sato.JJAP}. If the localized $d$ states of
impurities reside in the band gap of GaN- and ZnO-based DMS, the
double-exchange interaction, competing with the antiferromagnetic
super-exchange interaction, is responsible for the dominant ferromagnetism
\cite{Akai.PRL}. However, both double- and super-exchange interactions cannot
account for the long-range magnetic order at low doping concentrations of a few
percent. On the other hand, if the $d$ states of the impurity states are within
the valence band, such as in GaAs:Mn, it was confirmed both theoretically
\cite{Dietl.Science} and experimentally \cite{kitchen.NAT} that ferromagnetic
correlations of impurity ions are mediated by free carriers via a strong
$p$-$d$ hybridization exchange interaction. Additionally, some studies in the
opposite limit of strongly localized carriers, which are bound to impurity
bands, known as bound magnetic polarons have been proposed \cite{Durst.PRB}.
These theories have been used, to some degree of success, to explain the
ferromagnetism in 3$d$ or 4$f$ TM doped DMS. However, for the 2$p$~LE doped
DMS, the mechanism should be different since such materials have special
magnetic properties such as long-range FM coupling, high hole concentration,
and only $p$ orbitals contribute to the magnetism.

Here, we propose that alignment of magnetic moments in 2$p$ LE
doped DMS is achieved through a $p$-$d$ hybridization-like $p$-$p$ coupling
interaction between the impurity $p$ states and the host $p$ states at the top
of the valence band. This interaction follows essentially from quantum
mechanical level repulsion, which ``pushes" the minority state upward, crossing
the Fermi level [see Fig.~1(a)]. Consequently, the $p$ states
split into more stable threefold $t_2$ states which are either fully
occupied or completely empty. The symmetry and wave function ($p$-like $t_2$)
of the impurity 2$p$ state are similar to those of the top valence band of the
III-nitride and II-oxide which consists mostly of anion $p$ orbitals.
Therefore, a strong $p$-$p$ coupling interaction between the impurity state and
valence band state is allowed near the Fermi level. Substitution
of C for N in AlN or C/N for O in ZnO introduces impurity moments as well as
holes. Different from Mn ions in GaAs:Mn which polarizes spin of holes in
opposite direction, the spin density near each anion impurity in $2p$ LE doped
DMS tends to align parallel to the moment of the impurity ion under the $p$-$p$
interaction, as illustrated in Fig.~3(a) and Fig.~4(a).
The strong $p$-$p$ interaction leads to stronger coupling between impurity and carrier
spin orientations. Sufficiently densed spin-polarized carriers are able to
effectively mediate an indirect, long-range ferromagnetic coupling between
the 2$p$ LE dopants, as illustrated in Fig.~4(b). The spatially extended
$p$ states of the host and the impurity are able to extend the
$p$-$p$ interaction and spin alighment to a large range and thus to facilitate long-range
magnetic coupling between the impurities. This model gives a reasonable explanation to the
experimentally observed ferromagnetism in ZnO doped with a small amount of
carbons \cite{pan}. Therefore, it is the $p$-$p$ coupling interaction that determines the
electronic and magnetic properties of 2$p$ LE doped DMS, and free carriers
play an essential role in mediating the spin alighment in such DMS.

N has limited solubility in ZnO ($1\%-3\%$) \cite{fons, li01} and ZnO:N is
expected to be a weak ferromagnet. To increase the
N solubility and enhance ferromagnetism in ZnO:N, we propose a co-doping
approach. The conventional co-doping
concept was proposed by Yamamoto and Katayama-Yoshida \cite{katayama} to solve
the problem of doping asymmetry in ZnO. Two acceptors are combined with a donor or an
isovalent impurity to form an acceptor defect complex. By choosing the
co-doping element carefully, the acceptor defect complex can increase the
solubility of the desired dopants and reduce the acceptor defect level.
Following this concept, Be-N co-doped ZnO was demonstrated, both theoretically
and experimentally, to be a promising $p$-type ZnO  \cite{li, sanmyo}. The
concentration of N dopants was increased from 1\% to 7\% with 4\% Be co-doping
\cite{sanmyo}. Furthermore, the ionization energy of the N dopants was reduced
from 0.4 eV in ZnO:N \cite{park} to 0.12-0.22~eV in Be-N co-doped ZnO
\cite{li}.

Besides its high solubility in ZnO, Be was chosen as a candidate for co-doping
of ZnO:N in the present study because it is nonmagnetic and it does not have
its own $p$ electrons. Therefore it would not interfere with the magnetic
ordering in ZnO:N. Based on the experimental ratio of Be to N concentration
\cite{sanmyo} and the co-doping theory, a defect complex consisting of one Be
atom and two nitrogen atoms (Be+2N) is considered. Results of our
first-principles total-energy calculations indicate that the most stable
configuration of the Be+2N defect complex in ZnO is a N-Be-N structure in which
the Be atom, substituting a cation, is between two
satellite N atoms occupying the nearest-neighboring O sites, similar to
the Ga+2N complex in ZnO \cite{yan}. The calculated ionization energy of this
complex is 0.2 eV which is in good agreement with results of earlier DFT-LDA
calculations (0.12-0.22~eV) \cite{li}. More importantly, this value is smaller
than that of ZnO:N (0.4~eV) \cite{park}. Furthermore, our calculations show
that the co-doping with Be stabilizes the FM state within the same structure.
The FM state of the two N dopants in the Be+2N defect complex is energetically
favored over the AFM state by 54~meV which is larger than that of the same
configuration but without Be co-doping (7~meV). The enhancement of the FM
state in ZnO:Be+2N may be attributed to the decrease in the ionization energy
of ZnO:N and lowering of the acceptor level by co-doping. This leads to an
increase in carrier concentration and enhancement of the $p$-$p$ coupling
interaction. It is noted that both effects play a key role in the FM coupling
of the 2$p$~LE doped DMS.
Moreover, the ZnO:Be+2N can be expected to have a higher Curie temperature than
ZnO:N due to the enhanced $p$-$p$ coupling interaction in the
former\cite{Dietl.Science}.

In summary, 2$p$~LE doped II-oxide and III-nitride semiconductors were studied
and are proposed to be potential candidates for novel DMS materials, which
broadens the horizon of currently known magnetic systems. Based on results of
our first-principles calculation, we propose $p$-$p$ coupling interaction
as the origin of ferromagnetic coupling in such DMS. Furthermore, Be co-doped
ZnO:N was investigated as a possible approach to stabilize and enhance the FM
state of ZnO:N. It would be interesting to grow these materials experimentally
and to explore their applications.

\clearpage

\begin{figure}
\caption{(color online) Total DOS (a) and partial DOS (b) of N doped ZnO and
total DOS of C doped AlN (c) and C doped ZnO (d). One dopant is incorporated
into the 108-atom supercell which corresponds to a doping concentration of
1.85~at.\%. The Fermi energy is set to zero.}
\end{figure}

\begin{figure}
\caption{(color online) The spin-density (a) and charge-density (b) of N doped
ZnO. Most of the spin density is localized on the doped N atom and its twelve
second neighboring O atoms. The spin density around the nearest neighboring Zn
atoms which are not in the plane shown here is much smaller.}
\end{figure}

\begin{figure}
\caption{(color online) The spin density due to two N dopants seperated by
3.136 \AA.  (a) FM coupling and (b) AFM coupling. The FM state is most stable
than the AFM state. Blue and green isosurfaces correspond to up and down spin
densities, respectively.}
\end{figure}

\begin{figure}
\caption{(color online) Schematic diagram showing ferromagnetic coupling
between spins of doped N impurities. (a) The magnetic moment of N polarizes
carriers and and aligns the spins of the carriers in the same direction as that
of the N dopant. (b) If the carrier concentration is sufficiently high, it is
able to effectively mediate indirect ferromagnetic coupling among nearly all
doped N ions due to the long ranged $p$-$p$ interaction. Cation are not shown
in the diagram.}
\end{figure}

\clearpage
\begin{figure}
\includegraphics[height=0.80\textwidth]{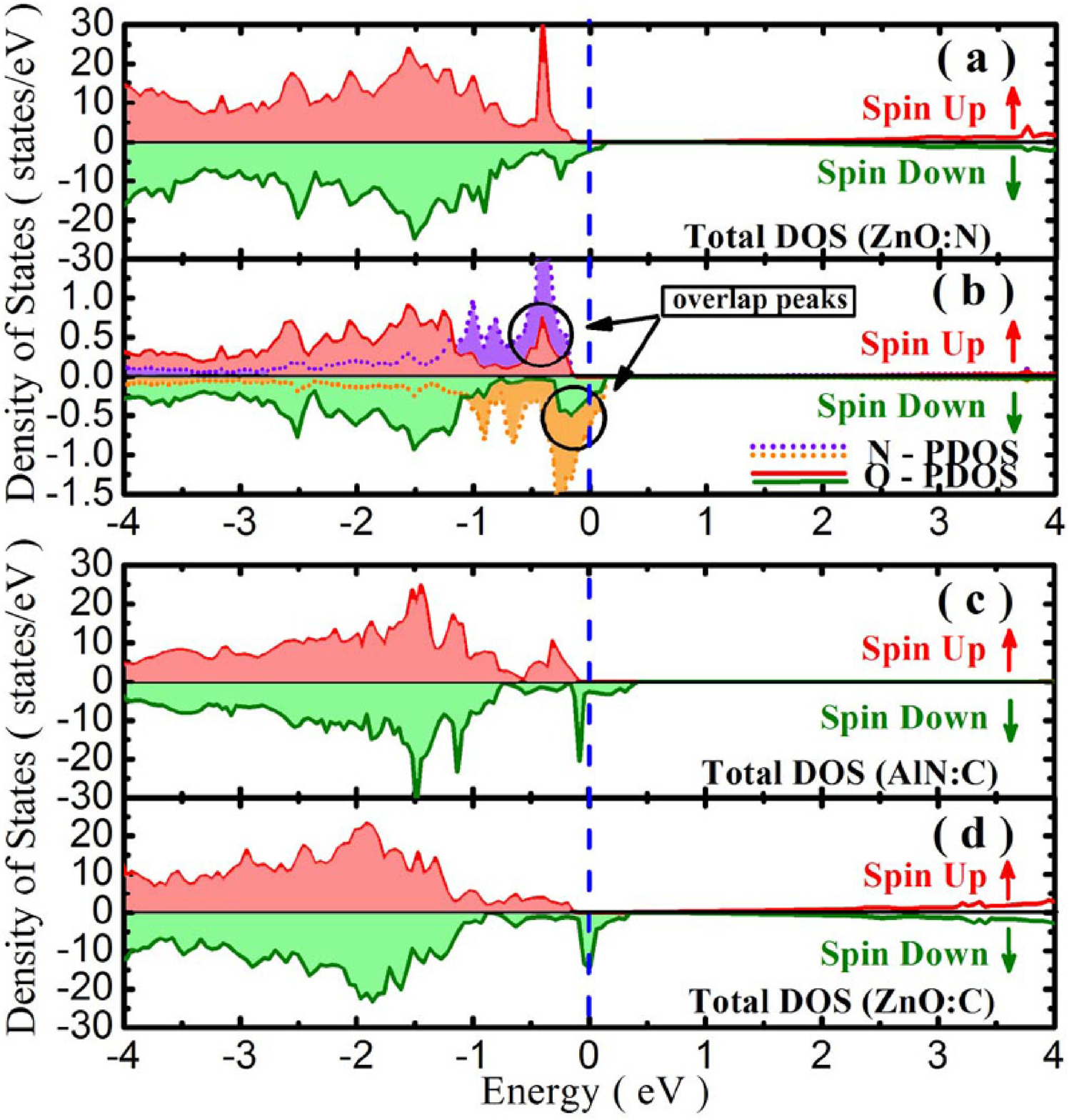}\\
{Fig.1 Shen et al.}
\end{figure}

\clearpage
\begin{figure}
\includegraphics[height=0.40\textwidth]{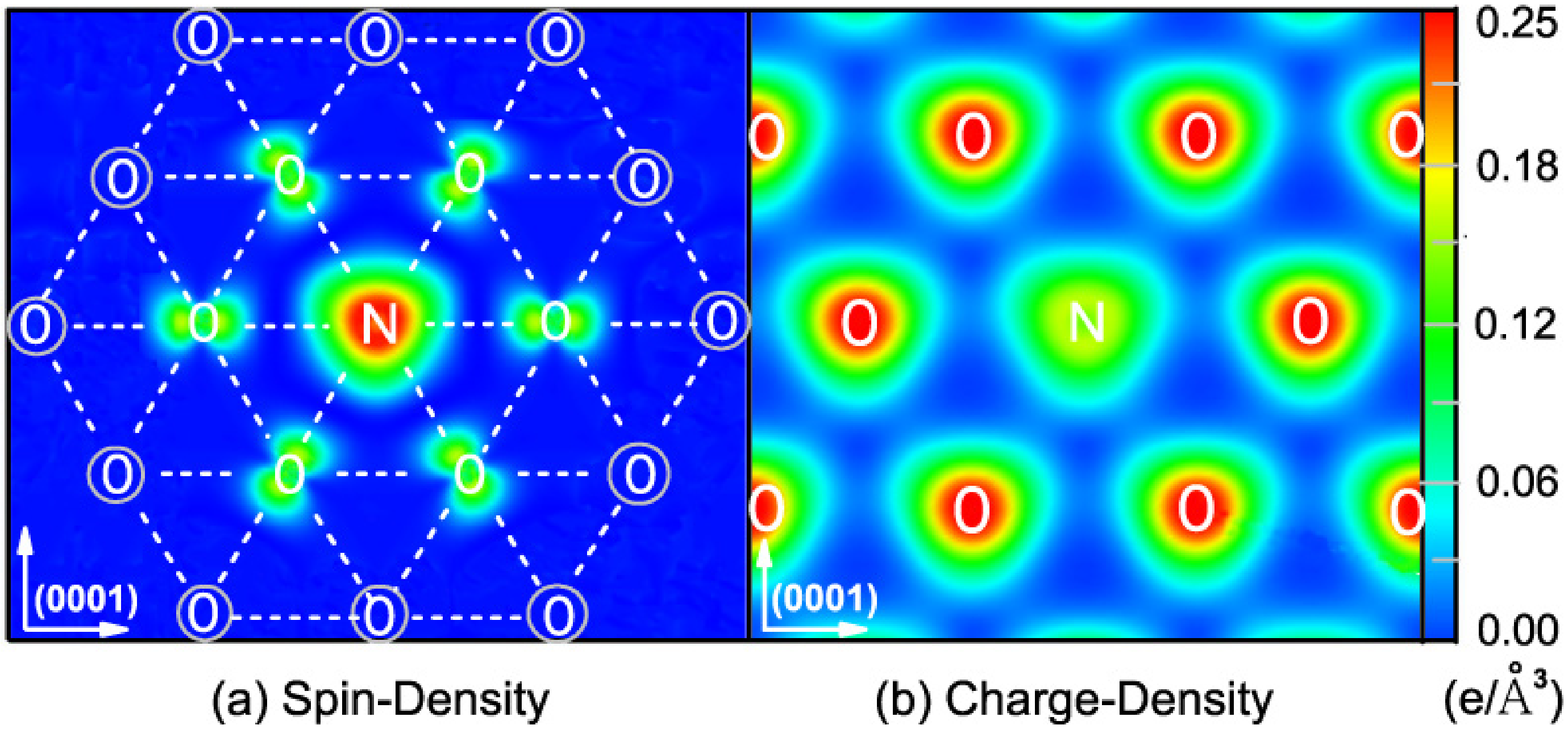}\\
{Fig.2 Shen et al.}
\end{figure}

\clearpage
\begin{figure}
\includegraphics[height=0.50\textwidth]{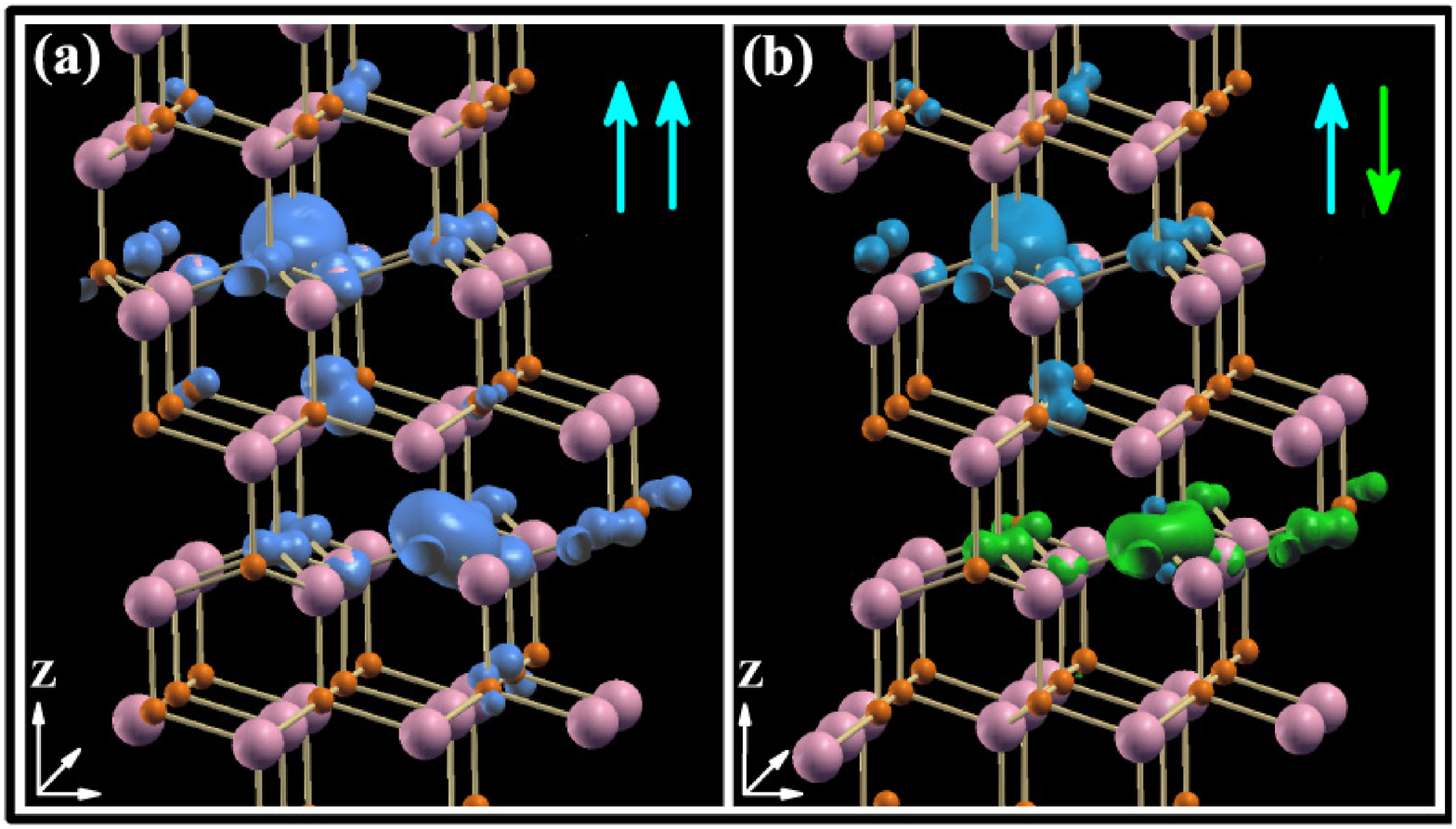}\\
{Fig.3 Shen et al.}
\end{figure}

\clearpage
\begin{figure}
\includegraphics[height=0.40\textwidth]{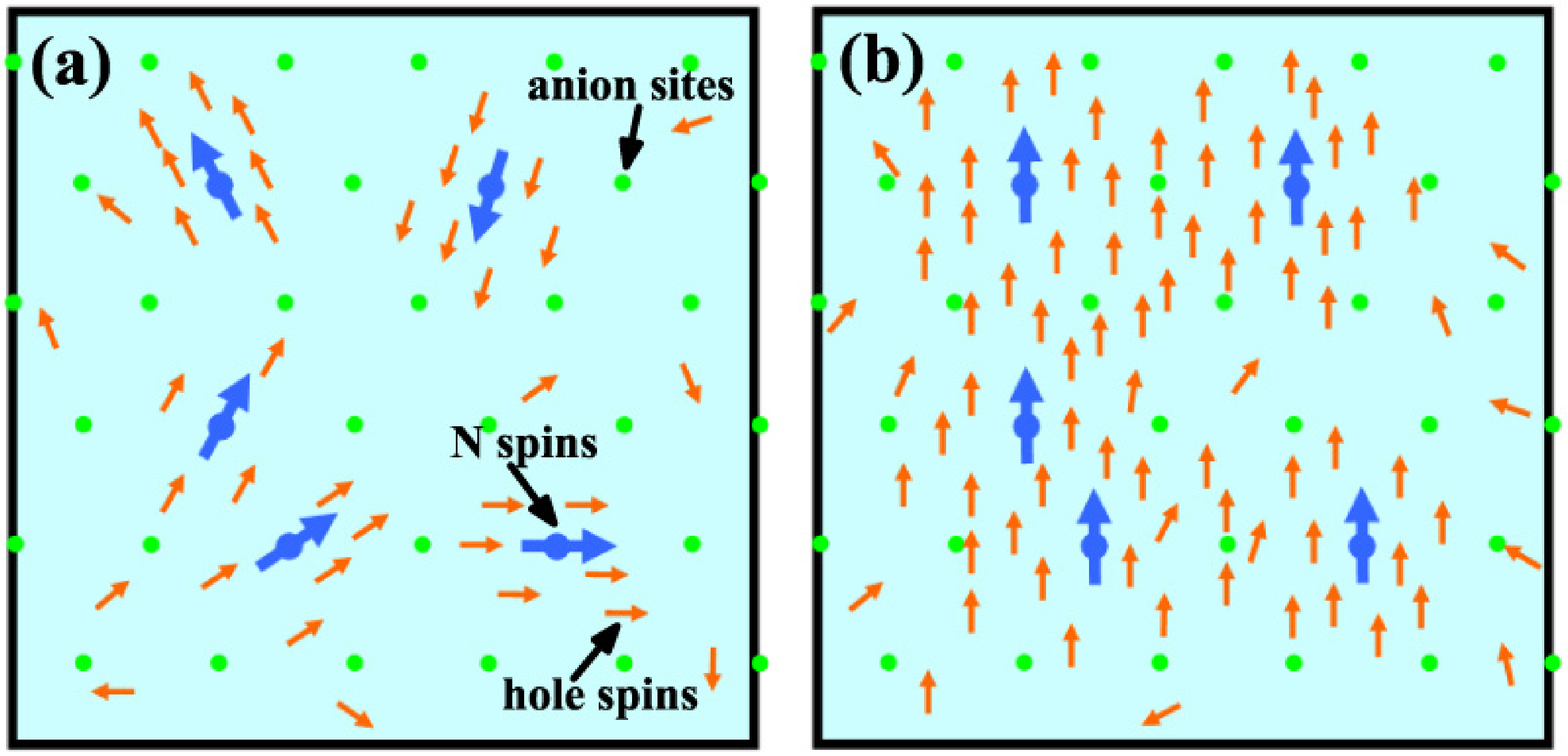}\\
{Fig.4 Shen et al.}
\end{figure}

\end{document}